# Goos-Hänchen and Imbert-Fedorov beam shifts: An overview

K. Y. Bliokh[a,b] and A. Aiello[c,d]

[a]*Applied Optics Group, School of Physics, National University of Ireland, Galway, Galway, Ireland*
[b]*A. Usikov Institute of Radiophysics and Electronics, Kharkov 61085, Ukraine*
[c]*Max Planck Institute for the Science of Light, Günter-Scharowsky-Straße 1/Bau 24, D-91058 Erlangen, Germany*
[d]*Institute for Optics, Information and Photonics, University Erlangen-Nürnberg, Staudtstraße 7/B2, 91058 Erlangen, Germany*

We consider reflection and transmission of polarized paraxial light beams at a plane dielectric interface. The field transformations taking into account a finite beam width are described based on the plane-wave representation and geometric rotations. Using geometrical-optics coordinate frames accompanying the beams, we construct an effective *Jones matrix* characterizing spatial-dispersion properties of the interface. This results in a unified self-consistent description of the *Goos-Hänchen* and *Imbert-Fedorov shifts* (the latter being also known as spin-Hall effect of light). Our description reveals intimate relation of the transverse Imbert-Fedorov shift to the *geometric phases* between constituent waves in the beam spectrum and to the *angular momentum* conservation for the whole beam. Both spatial and angular shifts are considered as well as their analogues for the higher-order vortex beams carrying intrinsic orbital angular momentum. We also give a brief overview of various extensions and generalizations of the basic beam-shift phenomena and related effects.

## 1. Introduction

Light reflection and refraction at a plane dielectric interface is one of the most basic optical processes known for ages and present in practically all optical systems. The interaction of a plane-wave with the interface is described by the well-known Snell's law and Fresnel formulas, which relate, respectively, the wave vectors and the polarization amplitudes of the incident and secondary waves [1]. This provides the geometrical-optics picture of light evolution. For a real optical beam which has a finite width (i.e., a distributed plane-wave spectrum) the situation becomes more complicated. It turns out that at the wavelength scale the reflected and transmitted beams do not exactly follow the geometrical-optics evolution. Neglecting shape deformations of the secondary beams, one can introduce *four* basic deviations from the geometrical-optics picture. With respect to the plane of incidence, these are in-plane and out-of-plane *spatial shifts* (i.e., lateral displacements) and similar *angular shifts* (i.e., deflections), see Fig. 1. The spatial and angular shifts can also be regarded as the *coordinate* and *momentum* shifts changing, respectively, positions and directions of propagation of the secondary beams. Following commonly accepted terminology, we will refer to these shifts as to the spatial and angular *Goos-Hänchen* (GH) and *Imbert-Fedorov* (IF) shifts. It turns out that all of these basic shifts can occur in a generic beam reflection or refraction, i.e., in almost every optical system. Currently, the GH and IF shifts are attracting rapidly growing attention caused by the development of nano-optics employing light evolution at subwavelength scales.

Originally, the spatial GH [2–5] and IF [6–8] shifts were discovered more than half-century ago for the total internal reflection. It was shown that both effects depend strongly on polarization of the incident beam. While the eigenmodes of the GH shift are TM (*p*) and TE (*s*) linearly polarized modes, the eigemodes of the IF effect are circularly (or elliptically) polarized waves.

Later, the angular GH shift was predicted for the case of partial reflection and transmission [9–11], which is also known as the *Fresnel filtering* [12,13]. This remarkable deviation from Newton's optics was recently measured experimentally [14]. The GH effects originate from the dispersion of the reflection or transmission coefficients as it was first shown by Artmann in 1948 [3]. For the past few decades, the spatial GH shift was studied in a variety of systems [15–28] embracing plasmonics, metamaterials, and quantum systems; now this effect is included in the classical electrodynamics textbooks [1].

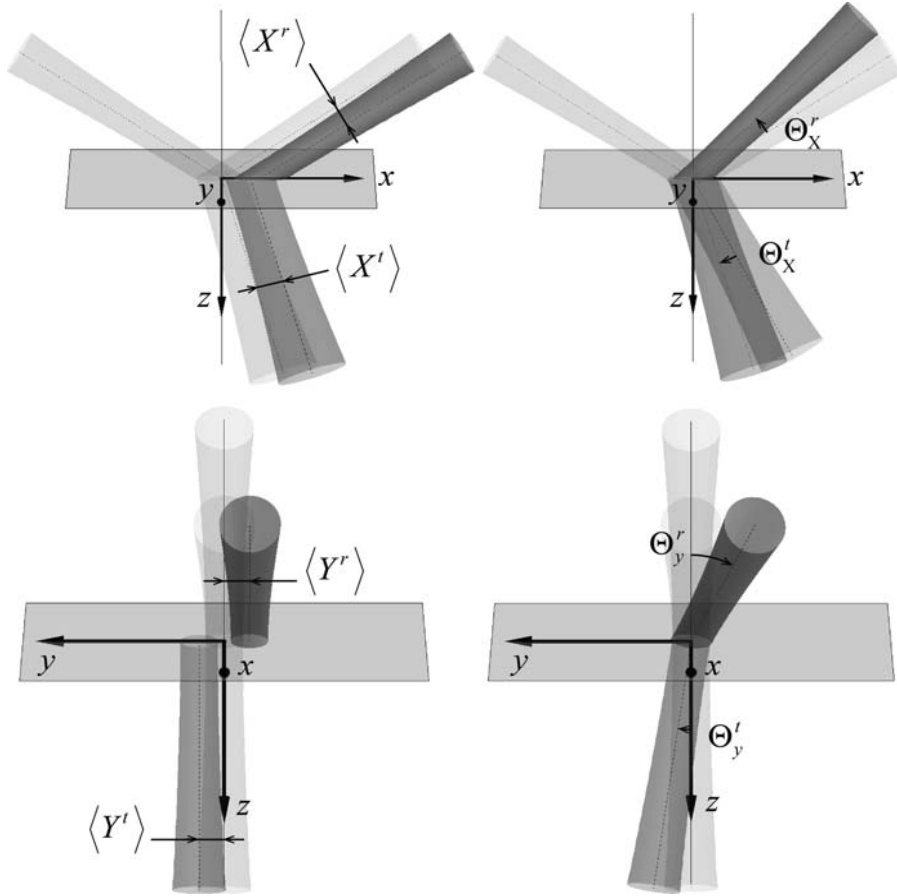

**Fig. 1.** Schematic picture of the beam shifts upon reflection and refraction at a plane interface ($z = 0$). The plane of incidence of the beam is $(x, z)$. In-plane (upper panels) and out-of-plane (lower panels), spatial (left panels) and angular (right panels) shifts of the reflected (*r*) and transmitted (*t*) beams. The displacements of the beam centroids, $\langle X^a \rangle$ and $\langle Y^a \rangle$, $a = r, t$, represent spatial Goos–Hänchen and Imbert–Fedorov shifts, respectively. The deflection angles $\Theta_x^a = \langle P_X^a \rangle / k^a$ and $\Theta_y^a = \langle P_y^a \rangle / k^a$ are associated with the angular (or momentum-space) Goos–Hänchen and Imbert–Fedorov shifts, respectively.

In contrast, the IF shift owes its origin to more deep and sophisticated physics, and the studies of the IF shift were notable for numerous controversies [29–53] even for the simplest case of the plane dielectric interface. Originally, Fedorov and Imbert explained this effect involving the Poynting energy flow arguments [6,8], whereas Schilling [7] was the first who derived in 1965 the adequate expression for the IF shift based on the plane-wave decomposition and interference inside the beams. In 1987, Fedoseyev and Player [40,41] showed that the IF shift is closely related to the balance and conservation of the *angular momentum* (AM) of light, including intrinsic spin AM associated with the circular polarization. In 1992, Liberman and Zel'dovich [43], introducing the notion of the *spin-orbit interaction* of light, re-derived the IF effect and Schilling's formula. In 2004, Onoda *et al.* [50] described the transverse shift as an example of the *spin-Hall effect* of light related to the *geometric Berry phase* and again re-derived the Schilling's formula based on the AM conservation. Finally, recently Bliokh & Bliokh gave a complete theoretical treatment to the IF shift, derived the exact expression improving the Schilling's formula, and predicted the angular IF shift [51,52]. This theory was verified experimentally by Hosten & Kwiat in 2008 [53]. Nowadays, it seems that all basic controversies are resolved, and the GH and IF shifts are studied within a unified approach considering beam transformations and deformations at the interface [54–56].

All the above descriptions of the GH and IF effects dealt with polarized beams of a Gaussian type. Because of the angular-momentum aspects of the IF effect, it is important to also consider higher-order beams carrying intrinsic *orbital* AM, i.e., the so-called *vortex beams* [57,58]. In 2001, Fedoseyev predicted the transverse shift induced by the vortex in such beams [59]. In 2006 the vortex-induced IF shift was detected experimentally for the reflected beam [60]. At the same time, the vortex-induced transverse shift of the transmitted beam was associated with conservation of the AM, *orbital Hall effect* and *orbit-orbit interaction* of light [61–63]. More recently, modification of all the four shifts (GH and IF, spatial and angular) for vortex beams were described theoretically [63] and verified experimentally for the reflected beam [64] (the orbital Hall effect for the transmitted beam is still to be observed). During the past few years, the rapidly growing interest to the spin-orbit interactions and Hall effects in optics resulted in a number of investigations of the transverse IF-type shifts for the beams carrying spin [65–74] and orbital [75–81] AM in various systems.

In the present paper, we give a self-consistent tutorial description of the GH and IF effects at a plane dielectric interface. In Section 2, we introduce the basic theoretical concepts involved in the subsequent analysis: field representations, rotations, geometric phases as well as the centroid, momentum, and AM of light beams. In Section 3, using geometrical rotations of constituent plane waves in the beam spectrum and standard Fresnel-Snell's equations, we describe the field transformations upon beam interaction with a dielectric interface. This enables us to construct an effective Jones matrix characterizing spatial-dispersion properties of the interface. The interface Jones matrix provides a unifying description of the IF and GH effects and illuminates the geometric-phase origin of the IF shift. Momentum and angular-momentum conservation underlying the shifts are also considered. In Section 4, we examine the beam shifts and underpinning conservation laws for the higher-order vortex beams carrying intrinsic orbital AM. Finally, Section 5 provides a brief overview of various extensions and generalizations of the basic GH, IF, and related beam-shift phenomena.

## 2. Basic concepts: Rotations, geometric phase, and angular momentum

### 2.1. Field representations and rotations

First, we consider a monochromatic light beam propagating along the *z*-axis in free space, which is characteried by the frequency $\omega$, the wave number $k = \omega$, and the complex electric field $\mathbf{E}(\mathbf{r})$. The actual real-valued field is $\boldsymbol{\mathcal{E}}(\mathbf{r},t) = \mathrm{Re}\left[\mathbf{E}(\mathbf{r})e^{-i\omega t}\right]$ and throughout the paper we use units $\hbar = c = 1$. The complex field can equivalently be characterized by its Fourier spectrum $\tilde{\mathbf{E}}(\mathbf{k})$:

$$\mathbf{E}(\mathbf{r}) \propto \int \tilde{\mathbf{E}}(\mathbf{k})e^{i\mathbf{k}\cdot\mathbf{r}}d^2\mathbf{k}_\perp . \qquad (2.1)$$

Here $d^2\mathbf{k}_\perp = dk_x dk_y$, $k_z(\mathbf{k}_\perp) = \sqrt{k^2 - k_x^2 - k_y^2}$, and we neglect the evanescent waves [55]. The plane-wave Fourier amplitudes $\tilde{\mathbf{E}}(\mathbf{k})$ determine the *momentum representation* of the field. We will also use the unit polarization vectors of the fields: $\mathbf{e} = \mathbf{E}/E$ and $\tilde{\mathbf{e}} = \tilde{\mathbf{E}}/\tilde{E}$.

From Maxwell equation $\nabla \cdot \mathbf{E} = 0$ in a homogeneous isotropic medium, it follows that

$$\mathbf{k} \cdot \tilde{\mathbf{E}} = 0. \tag{2.2}$$

This is the *transversality condition* which compels the electric field of a plane wave, $\tilde{\mathbf{E}}$, to be orthogonal to its wave vector $\mathbf{k}$. This constraint couples *polarization* and *momentum* of light, which is a typical feature of the spin-orbit interaction. The condition (2.2) has a natural geometrical representation: the wave electric field must be tangent to the surface of the *unit sphere of directions*, $S^2 = \{\boldsymbol{\kappa}\}$, $\boldsymbol{\kappa} = \mathbf{k}/k$, in momentum $\mathbf{k}$-space, Fig. 2.

Alongside with the 'absolute' field vectors $\mathbf{E}$ and $\tilde{\mathbf{E}}$, which are independent of the choice of the coordinate frame, we introduce 'vectors of components', $\left|\mathbf{E}\right)_A$ and $\left|\tilde{\mathbf{E}}\right)_A$, i.e., field components in the given coordinate frame $A$. For instance, in the global Cartesian frame $(x, y, z)$, the *laboratory frame*, we have

$$\mathbf{E} = E_x \mathbf{u}_x + E_y \mathbf{u}_y + E_z \mathbf{u}_z, \quad \left|\mathbf{E}\right)_L = \begin{pmatrix} E_x \\ E_y \\ E_z \end{pmatrix}, \tag{2.3}$$

where $\mathbf{u}_\alpha$ stands for the unit basis vector of the corresponding $\alpha$-axis. We will also use the transverse vector components considered with respect to the $z$-axis of the frame in use and denoted by the "$\perp$" subscript: $\left|\mathbf{E}\right)_{\perp L} = \left(E_x, E_y\right)^T$, $\mathbf{r}_\perp = x\mathbf{u}_x + y\mathbf{u}_y$, etc.

Components of the same vector in different bases are related by a *unitary transformation* which links the basic vectors of the frames:

$$\begin{pmatrix} E'_x \\ E'_y \\ E'_z \end{pmatrix} = \hat{U} \begin{pmatrix} E_x \\ E_y \\ E_z \end{pmatrix}, \quad \begin{pmatrix} \mathbf{u}'_x \\ \mathbf{u}'_y \\ \mathbf{u}'_z \end{pmatrix} = \hat{U} \begin{pmatrix} \mathbf{u}_x \\ \mathbf{u}_y \\ \mathbf{u}_z \end{pmatrix}. \tag{2.4}$$

For instance, in the *'laboratory circular basis'* $\left(\mathbf{u}_0^+, \mathbf{u}_0^-, \mathbf{u}_z\right)$ of the circular polarizations in the $(x, y)$ plane, with $\mathbf{u}_0^\pm = \left(\mathbf{u}_x \pm i\mathbf{u}_y\right)/\sqrt{2}$, one has $E_0^\pm = \left(E_x \mp iE_y\right)/\sqrt{2}$, i.e.,

$$\left|\mathbf{E}\right)_C = \hat{V}\left|\mathbf{E}\right)_L, \quad \hat{V} = \frac{1}{\sqrt{2}} \begin{pmatrix} 1 & -i & 0 \\ 1 & i & 0 \\ 0 & 0 & \sqrt{2} \end{pmatrix}. \tag{2.5}$$

For the wave propagating along the $z$-axis, $\mathbf{k} = k\mathbf{u}_z$, the polarization vectors of right- and left-circularly polarized fields ($\sigma = \pm 1$) are given by:

$$\left|\tilde{\mathbf{E}}^\sigma\right)_C \propto \left|\mathbf{e}^\sigma\right), \quad \left|\mathbf{e}^+\right) = (1, 0, 0)^T, \quad \left|\mathbf{e}^-\right) = (0, 1, 0)^T. \tag{2.6}$$

Here $\sigma$ represents the helicity quantum number, which determines the spin states of photons. Polarizations vectors (2.6) are eigenvectors of the diagonal *helicity operator* $\hat{\sigma}$:

$$\hat{\sigma} = \text{diag}(1, -1, 0), \quad \hat{\sigma}\left|\mathbf{e}^\sigma\right) = \sigma\left|\mathbf{e}^\sigma\right), \quad \sigma = \pm 1. \tag{2.7}$$

The third eigenvector of $\hat{\sigma}$, $\left|\mathbf{e}_z\right) = (0, 0, 1)^T$, with eigenvalue $\sigma = 0$, corresponds to the longitudinal polarization prohibited by Eq. (2.2).

Since the transversality condition (2.2) attaches the generic wave electric field $\tilde{\mathbf{E}}$ to the surface of the $\boldsymbol{\kappa}$-sphere, it is natural to describe the polarization evolution in the *spherical coordinates* $(\theta, \phi, k)$ in momentum space (see Fig. 2). Their basic vectors are

$$\mathbf{u}_\theta(\mathbf{k}) = \mathbf{u}_\phi(\mathbf{k}) \times \boldsymbol{\kappa}, \quad \mathbf{u}_\phi(\mathbf{k}) = \frac{\mathbf{u}_z \times \boldsymbol{\kappa}}{|\mathbf{u}_z \times \boldsymbol{\kappa}|}, \quad \mathbf{u}_k(\mathbf{k}) = \boldsymbol{\kappa} . \tag{2.8}$$

Transformation between the laboratory and spherical frames can be represented by the product of two rotations by the angles $\theta$ and $\phi$, $\hat{U}_S = \hat{R}_y(\theta)\hat{R}_z(\phi)$:

$$\left|\tilde{\mathbf{E}}\right\rangle_S = \hat{U}_S(\mathbf{k})\left|\tilde{\mathbf{E}}\right\rangle_L, \quad \hat{U}_S = \begin{pmatrix} \cos\theta\cos\phi & \cos\theta\sin\phi & -\sin\theta \\ -\sin\phi & \cos\phi & 0 \\ \sin\theta\cos\phi & \sin\theta\sin\phi & \cos\theta \end{pmatrix}. \tag{2.9}$$

Here we introduced the operators of SO(3) rotations: $\hat{R}_\alpha(\delta) = \exp(i\delta\hat{S}_\alpha)$, $\alpha = x, y, z$, where $(\hat{S}_\alpha)_{ij} = -i\varepsilon_{\alpha ij}$ ($\varepsilon_{\alpha ij}$ is the Levi-Civita symbol) are the generators of rotations, i.e., spin-1 matrices. In spherical coordinates, the transversality condition (2.2) becomes particularly simple: $\tilde{E}_k = 0$, so that the field is two-component: $\left|\tilde{\mathbf{E}}\right\rangle_S = \left(\tilde{E}_\theta, \tilde{E}_\phi, 0\right)^T$.

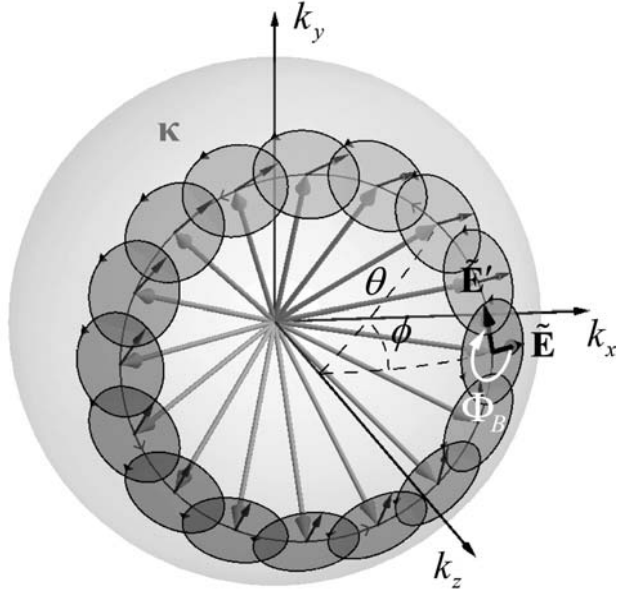

**Fig. 2.** The unit sphere $S^2 = \{\boldsymbol{\kappa}\}$ in the momentum space represents the directions of propagation of plane waves specified by spherical angles $(\theta, \phi)$. The electric field $\tilde{\mathbf{E}}$ is tangent to this sphere. Parallel transport of the field along the contour $C$ with a fixed polar angle $\theta$ brings about rotation $\tilde{\mathbf{E}} \rightarrow \tilde{\mathbf{E}}'$ by the angle numerically equal to the Berry phase $\Phi_B(C) = -2\pi\cos\theta$, Eqs. (2.13)–(2.15). Correspondingly, the circularly polarized field components in the spherical coordinates, Eq. (2.10), acquire the phase factors $\exp(-i\sigma\Phi_B)$, Eq. (2.13).

Spherical coordinates provide a description of a plane wave in the field spectrum in the local basis attached to its $\mathbf{k}$-vector. Akin to the laboratory circular-polarization basis $\mathbf{u}_0^\sigma$, the circular polarizations defined with respect to the spherical vectors form the *helicity basis* [82,83]:

$$\mathbf{u}^\sigma = \frac{\mathbf{u}_\theta + i\sigma\mathbf{u}_\phi}{\sqrt{2}}, \quad \sigma = \pm 1. \tag{2.10}$$

This basis has the polar singularity at $\theta = 0$. In the paraxial limit $\theta \to 0$ (propagation along the $z$-axis), the basic vectors $\mathbf{u}^\sigma$ tend to the laboratory circular-polarization basis with an additional azimuthal rotation: $\mathbf{u}^\sigma \to \mathbf{u}_0^\sigma e^{-i\sigma\phi}$.[1] Transformation from the laboratory circular frame (2.5) to the helicity basis (2.10) is given by the unitary matrix $\hat{U}_H(\mathbf{k}) = \hat{V}\hat{R}_y(\theta)\hat{R}_z(\phi)\hat{V}^\dagger$.

## 2.2. Geometric phase

Any polarization state is defined up to a common phase. In the case of circular polarization, the phase can be induced by a *rotation* of the coordinate frame about the $\mathbf{k}$-vector. For instance, if a plane wave propagates along the $z$-axis, rotation of the coordinates $(x, y)$ by the angle $-\Phi$, $|\mathbf{E}'\rangle_L = \hat{R}_z(-\Phi)|\mathbf{E}\rangle_L$, induces the *geometric phase* $\sigma\Phi$ in the circular field components:

$$\mathbf{u}_0^{\sigma\prime} = e^{i\sigma\Phi}\mathbf{u}_0^\sigma, \quad |\mathbf{E}'\rangle_C = \exp\left(-i\hat{\sigma}\Phi\right)|\mathbf{E}\rangle_C, \quad e^{-i\hat{\sigma}\Phi} = \begin{pmatrix} e^{-i\Phi} & 0 & 0 \\ 0 & e^{i\Phi} & 0 \\ 0 & 0 & 1 \end{pmatrix}. \tag{2.11}$$

This is the simplest 2D example of the *spin-rotation coupling* [84–87].

In the general 3D case, polarizations of different waves are defined in the *different* planes orthogonal to the wave vectors $\mathbf{k}$, i.e., tangent to the surface of the $\boldsymbol{\kappa}$-sphere, Fig. 2. Hence, to compare the phases and polarizations of plane waves with different $\mathbf{k}$-vectors, one has to transport their electric fields over the surface of sphere in order to bring them to the same plane. The geometric *parallel transport* of vectors tangent to the sphere is defined in such a way that the vector does not experience *local* rotation about the normal $\boldsymbol{\kappa}$-vector. However, the parallel-transport frame cannot be defined *globally* on the sphere, and *any globally-defined coordinate system on the sphere inevitably experiences local rotations of its axes about the $\boldsymbol{\kappa}$-vector and induces geometric phases for circularly polarized modes*. In particular, for the spherical coordinates (2.8) and helicity basic vectors (2.10), the corresponding differential angle of rotation and the geometric phase reads [88–91]:

$$-\sigma d\Phi = i\mathbf{u}^{\sigma*} \cdot d\mathbf{u}^\sigma = i\sum_j \left[\mathbf{u}^{\sigma*}(\mathbf{k}) \cdot \left(\frac{\partial}{\partial k_j}\right)\mathbf{u}^\sigma(\mathbf{k})\right]dk_j, \tag{2.12}$$

For the evolution of the field along a contour $C$ between $\mathbf{k}$ and $\mathbf{k}'$ in momentum space (owing to the transversality, it can always be projected onto the $\boldsymbol{\kappa}$-sphere), integration of Eq. (2.12) results in the accumulation of the geometric phase (2.11):

$$|\tilde{\mathbf{E}}'\rangle_H = \exp\left(-i\hat{\sigma}\Phi_B\right)|\tilde{\mathbf{E}}\rangle_H, \quad \Phi_B(C) = \int_C \mathbf{A}_B(\mathbf{k}) \cdot d\mathbf{k} = -\int\cos\theta\, d\phi, \tag{2.13}$$

$$\mathbf{A}_B = -i\sigma^{-1}\mathbf{u}^{\sigma*} \cdot \left(\frac{\partial}{\partial\mathbf{k}}\right)\mathbf{u}^\sigma = -k^{-1}\cot\theta\,\mathbf{u}_\phi. \tag{2.14}$$

Here $\Phi_B$ is the *Berry geometric phase* described by the momentum-space contour integral of the *Berry connection* $\mathbf{A}_B(\mathbf{k})$ which was calculated from Eqs. (2.8), (2.10), and (2.12). It follows from Eq. (2.13) that the Berry phase difference between the waves with a fixed polar angle $\theta = \text{const}$ distributed in the range of azimuthal angles $(0, \phi)$ is equal to

$$\Phi_B = -\phi\cos\theta. \tag{2.13'}$$

---

[1] This additional rotation is related to the polar singularity and rotation of the spherical coordinates. It can be removed in the helicity basis by gauge transformation $\mathbf{u}^\sigma \to \mathbf{u}^\sigma e^{i\sigma\phi}$, which provides non-singular transition to the paraxial limit $\theta \to 0$ [82,83]. However, in our consideration there is no need for that.

An example of the Berry phase $\Phi_B = -2\pi\cos\theta$ caused by the parallel transport of the field along a closed contour of evolution $C = \{\theta = \text{const}, \phi \in (0, 2\pi)\}$ is shown in <span style="color:orange">Fig. 2</span>.

Berry phase (2.13) has been extensively studied in various optical systems (for reviews, see [89–93]). It is associated with the variations of the **k**-vector and becomes significant in globally *nonparaxial* optical systems involving 3D distributions of the wave vectors. One can distinguish the two typical situations: (i) successive nonplanar variations of the direction of propagation of a polarized wave (in multiple reflections, refractions, curved optical fibers, etc.) [89–93]; and (ii) simultaneous interference involving multiple polarized waves with different **k**-vectors (e.g., tightly focused or scattered nonparaxial fields) [83,87]. As we will see, the Berry phase of the second type underlies the spin Hall effect of light, i.e., the IF shifts of polarized beams.

The Berry phase also allows simple dynamical interpretations. Upon the evolution of the **k**-vector, the electric field $\tilde{\mathbf{E}}(\mathbf{k})$ possesses a sort of *inertia* and remains locally non-rotating about **k**. Observation in a globally-defined spherical frame produces a *Coriolis effect* caused by local rotations of the basis. This is expressed in the simple 'dynamical' formulae for the Berry phase [87,91,94]:

$$\sigma\Phi_B = -\sigma\int\boldsymbol{\kappa}\cdot\boldsymbol{\Omega}_\tau\,d\tau. \tag{2.15}$$

Here $\tau$ is a parameter underlying the evolution of waves in the momentum space (this can be time, a coordinate in real space, etc.) and $\boldsymbol{\Omega}_\tau$ is the angular velocity of rotation of the coordinate frame defined with respect to this parameter $\tau$. In particular, waves with fixed polar angle $\theta = \text{const}$ but different azimuthal angles $\phi$ on the $\boldsymbol{\kappa}$-sphere are related via rotation by the angle $\phi$ about the $z$-axis, i.e., $\boldsymbol{\Omega}_\phi = \mathbf{u}_z$ and $\Phi_B = -\int\boldsymbol{\kappa}\cdot\mathbf{u}_z\,d\phi = -\int\cos\theta\,d\phi$ [87,91,94], cf. Eq. (2.13). Equation (2.15) represents the Berry phase as a result of the spin-rotation coupling between the spin AM $\sigma\boldsymbol{\kappa}$ and the rotation $\boldsymbol{\Omega}_\tau$. In this spirit, it is completely analogous to the Coriolis effect [84–87] and the rotational Doppler effect for waves carrying intrinsic AM [95–98].

### 2.3. Coordinate, momentum, and angular momentum

While the Berry phase describes the spin-orbit interaction phenomena on the level of geometric interference of *planes waves* forming the field, the coordinate and AM represent integral dynamical characteristics of the *localized wave packets* or *beams*. They are naturally described within a quantum-like operator formalism.

The operators for the coordinate, momentum, and energy of a photon are written in the momentum representation as:

$$\hat{\mathbf{r}} = i\frac{\partial}{\partial\mathbf{k}}, \quad \hat{\mathbf{p}} = \mathbf{k}, \quad \hat{w} = \omega, \tag{2.16}$$

whereas the canonical orbital and spin AM operators are known to be [99]

$$\hat{\mathbf{L}} = -i\left(\mathbf{k}\times\frac{\partial}{\partial\mathbf{k}}\right), \quad \left(\hat{S}_\alpha\right)_{ij} = -i\varepsilon_{\alpha ij}. \tag{2.17}$$

The orbital and spin AM, $\hat{\mathbf{L}}$ and $\hat{\mathbf{S}}$, represent differential and matrix operators that act in Hilbert-momentum and vector-polarization spaces, respectively.[2] The spin operator $\hat{\mathbf{S}}$ consists of the generators of SO(3) rotations that act on the components of the field in the laboratory frame, $|\tilde{\mathbf{E}}\rangle_L$.

---

[2] There are some fundamental difficulties in using canonical operators (2.16) and (2.17) for generic nonparaxial fields [83,99,100], but here we restrict ourselves to the paraxial approximation, which is free of such problems.

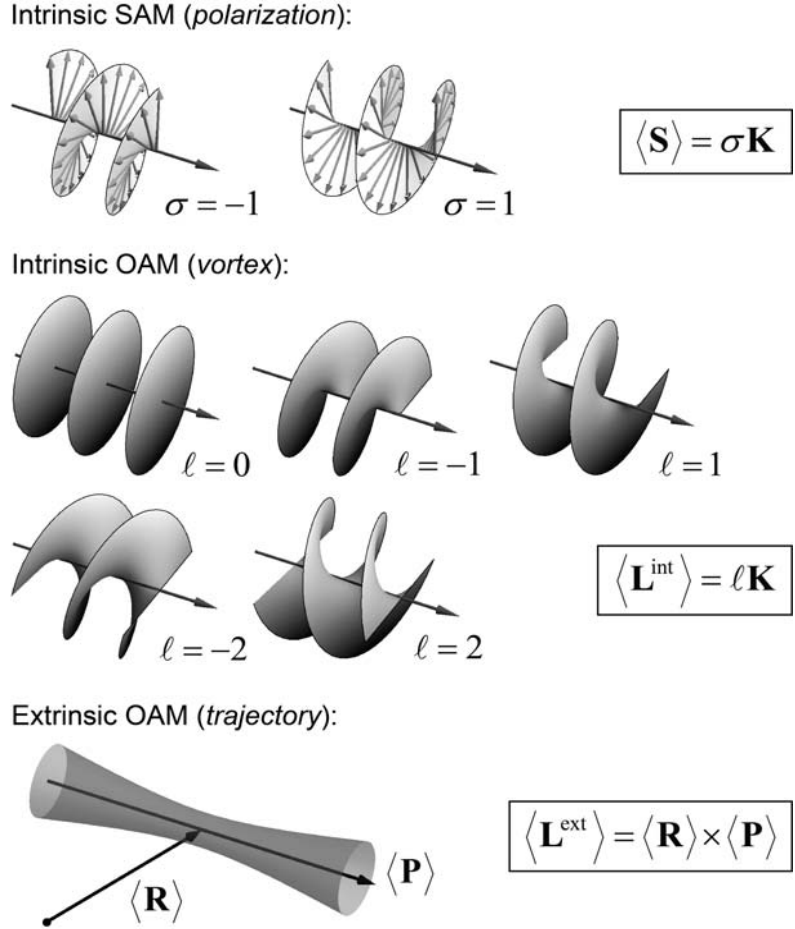

**Fig. 3.** The three types of optical AM for paraxial light: (i) spin, (ii) intrinsic orbital, and (iii) extrinsic orbital AM. They are associated with circular polarization, an optical vortex, and the motion of the field centroid, respectively. The quantum number $\sigma = \pm 1$ indicates the helicity of the right-hand and left-hand circular polarizations, while $\ell = 0, \pm 1, \pm 2, ...$ indicates the charge of optical vortex. The equations show the normalized values of the AM per photon (we use units $\hbar = 1$), where $\langle \mathbf{R} \rangle$ stands for the radius-vector of the centroid of the optical field, $\langle \mathbf{P} \rangle$ is the mean momentum (wave-vector), and $\mathbf{K} = \langle \mathbf{P} \rangle / \langle P \rangle$ is the direction of propagation, see Eqs. (2.20) and (2.22).

Amounts of energy, momentum, and AM carried by a light beam propagating along the $z$-axis are characterized by the linear densities of these quantities per unit $z$-length [58,101]. In this manner, the normalized linear densities of the energy, $\langle W \rangle$, momentum, $\langle \mathbf{P} \rangle$, orbital AM, $\langle \mathbf{L} \rangle$, and spin AM, $\langle \mathbf{S} \rangle$, as well as the coordinates of the field centroid, $\langle \mathbf{R} \rangle$, can be calculated as *expectation values* of the operators (2.16) and (2.17):

$$\langle \mathbf{R} \rangle = N^{-1} \left\langle \tilde{\mathbf{E}} \middle| \hat{\mathbf{r}} \middle| \tilde{\mathbf{E}} \right\rangle_L, \quad \langle \mathbf{P} \rangle = N^{-1} \left\langle \tilde{\mathbf{E}} \middle| \mathbf{k} \middle| \tilde{\mathbf{E}} \right\rangle_L, \quad \langle W \rangle = N^{-1} \left\langle \tilde{\mathbf{E}} \middle| \hat{w} \middle| \tilde{\mathbf{E}} \right\rangle_L. \tag{2.18}$$

$$\langle \mathbf{L} \rangle = N^{-1} \left\langle \tilde{\mathbf{E}} \middle| \hat{\mathbf{L}} \middle| \tilde{\mathbf{E}} \right\rangle_L, \quad \langle \mathbf{S} \rangle = N^{-1} \left\langle \tilde{\mathbf{E}} \middle| \hat{\mathbf{S}} \middle| \tilde{\mathbf{E}} \right\rangle_L. \tag{2.19}$$

Here $N = \left\langle \tilde{\mathbf{E}} \middle| \tilde{\mathbf{E}} \right\rangle_L$ is the norm (the number of photons per unit $z$-length), whereas the state vector $\left| \tilde{\mathbf{E}} \right\rangle_L = \left| \tilde{\mathbf{E}}(\mathbf{k}_\perp) \right\rangle_L \exp\left[ ik_z(\mathbf{k}_\perp) z \right]$ is proportional to the field polarization vector $\left| \tilde{\mathbf{E}} \right\rangle_L$ but is also considered as a vector in the Hilbert space [55,83]. Correspondingly, the 'bra-ket' inner product

implies the scalar product of complex polarizations together with the integration over the momentum space: $\left\langle \tilde{\mathbf{E}} \middle| \tilde{\mathbf{E}}' \right\rangle_L = \int \left( \tilde{\mathbf{E}} \middle| \tilde{\mathbf{E}}' \right)_L d^2 \mathbf{k}_\perp$.

While the spin AM $\langle \mathbf{S} \rangle$ is purely intrinsic (i.e., independent of the origin), the orbital AM of light can be divided into *extrinsic* (origin-dependent) and *intrinsic* contributions, $\left\langle \mathbf{L}^{\text{ext}} \right\rangle$ and $\left\langle \mathbf{L}^{\text{int}} \right\rangle$ [83,102]. The extrinsic contribution is determined by the motion of the centroid of the field and is given by the 'mechanical' cross-product of the central coordinate and momentum [40,41,50–52,61–63,83]:

$$\left\langle \mathbf{L}^{\text{ext}} \right\rangle = \langle \mathbf{R} \rangle \times \langle \mathbf{P} \rangle, \quad \left\langle \mathbf{L}^{\text{int}} \right\rangle = \langle \mathbf{L} \rangle - \left\langle \mathbf{L}^{\text{ext}} \right\rangle. \tag{2.20}$$

Thus, there are three types of the AM of light: (i) spin, (ii) intrinsic orbital, and (iii) extrinsic orbital AM. For the locally-paraxial fields they are associated, respectively, with circular polarization, optical vortices, and transverse beam shifts, Fig. 3. If the medium possesses rotational symmetry about certain axis, the corresponding component of the total AM, $\langle \mathbf{L} \rangle + \langle \mathbf{S} \rangle = \left\langle \mathbf{L}^{\text{int}} \right\rangle + \left\langle \mathbf{L}^{\text{ext}} \right\rangle + \langle \mathbf{S} \rangle$, is conserved upon evolution of light. However, mutual conversion between different parts of the AM is possible which signals the *spin-orbit* (or *orbit-orbit*) *interaction* of light. In a similar way, if the medium is stationary and possesses translational symmetry about certain axis, the energy $\langle W \rangle$ and corresponding component of the momentum $\langle \mathbf{P} \rangle$ must be conserved upon light evolution.

Considering paraxial beams propagating along the $z$-axis, $\theta \ll 1$, we neglect small longitudinal $z$-components of the field and deal with the transverse $(x, y)$-components denoted by $\left| \tilde{\mathbf{E}} \right\rangle_\perp$. As an example, let us consider paraxial circularly-polarized Laguerre–Gaussian vortex beams with the azimuthal quantum number $\ell = 0, \pm 1, \pm 2, ...$ and radial quantum number $p = 0$ [58]. The electric field in the laboratory basis of circular polarizations can be written as:

$$\left| \tilde{\mathbf{E}}_\ell^\sigma (\mathbf{k}) \right\rangle_{\perp C} = \left| \mathbf{e}^\sigma \right\rangle_\perp \left| \tilde{E}_\ell (\mathbf{k}) \right\rangle, \quad \left| \tilde{E}_\ell (\mathbf{k}) \right\rangle = \theta^{|\ell|} G(\theta) e^{i\ell\phi + ik(1 - \theta^2/2)z}, \tag{2.21}$$

where $G(\theta) \propto \exp\left[ -(kw_0)^2 \theta^2 / 4 \right]$ is the Gaussian envelope function with $w_0 \gg k^{-1}$ being the beam waist. The azimuthal phase factor $\exp(i\ell\phi)$ in Eq. (2.21) represents the *optical vortex* of charge $\ell$. Note that the spin and orbital degrees of freedom are factorized here into the constant polarization vector $\left| \mathbf{e}^\sigma \right\rangle$ in the real space and scalar field $\left| \tilde{E}_\ell (\mathbf{k}) \right\rangle$ as a vector in the Hilbert space. Substituting field (2.21) into Eqs. (2.18) and (2.19), we obtain the expectation values:

$$\langle \mathbf{R} \rangle = z\mathbf{K}, \quad \langle \mathbf{P} \rangle = k\mathbf{K}, \quad \langle W \rangle = \omega, \quad \langle \mathbf{L} \rangle = \ell\mathbf{K}, \quad \langle \mathbf{S} \rangle = \sigma\mathbf{K}. \tag{2.22}$$

where $\mathbf{K} = \mathbf{u}_z$ is the direction of propagation of the beam and we took into account that the spin operator in the basis of circular polarizations (2.5) is $\left( \hat{\mathbf{S}} \right)_C = \hat{V} \hat{\mathbf{S}} \hat{V}^\dagger$. Equations (2.22) represent well-known results for the intrinsic orbital and spin AM of the paraxial vortex beams [57,58] (see Fig. 3). It is easy to see that beams (2.21) are eigenmodes of the $\hat{L}_z$ and $\left( \hat{S}_z \right)_C$ operators with the discrete eigenvalues $\ell$ and $\sigma$. Indeed, these operators read

$$\hat{L}_z = -i \frac{\partial}{\partial \phi}, \quad \left( \hat{S}_z \right)_C = \hat{V} \hat{S}_z \hat{V}^\dagger = \hat{\sigma}, \tag{2.23}$$

so that vortices $\exp(i\ell\phi)$ and polarizations $\left| \mathbf{e}^\sigma \right\rangle$, Eqs. (2.6) and (2.7), represent their eigenmodes. Obviously, the extrinsic AM (2.20) vanishes for the beam (2.21) and (2.22): $\left\langle \mathbf{L}^{\text{ext}} \right\rangle = \mathbf{0}$. However, mutually orthogonal transverse shift and tilt of the beam would generate nonzero extrinsic AM $\left\langle \mathbf{L}^{\text{ext}} \right\rangle = \left\langle \mathbf{R} \right\rangle \times \left\langle \mathbf{P} \right\rangle \neq 0$.

Note that transverse part of the helicity operator (2.7) represents the Pauli matrix, $\hat{\sigma}_{\perp} = \text{diag}\left(1,-1\right) = \hat{\sigma}_3$, with the corresponding eigenvectors of circular polarizations: $\left|\mathbf{e}^+\right\rangle_{\perp} = \left(1,0\right)^T$ and $\left|\mathbf{e}^-\right\rangle_{\perp} = \left(0,1\right)^T$. An arbitrary uniform polarization state of the paraxial field is described by the complex unit *Jones vector* $\left|\mathbf{e}\right\rangle_{\perp C} = \left(e^+, e^-\right)^T$, $\left|e^+\right|^2 + \left|e^-\right|^2 = 1$, obeying $\text{SU}\left(2\right)$ symmetry of a two-level system. For the beam with such polarization, the helicity $\sigma$ in Eqs. (2.22) is replaced by the mean helicity $\bar{\sigma} = \left(\mathbf{e}\middle|\hat{\sigma}_3\middle|\mathbf{e}\right)_{\perp C} = \left|e^+\right|^2 - \left|e^-\right|^2$. Thus, the Pauli matrix $\hat{\sigma}_3$ underpins the degree of the circular polarization responsible for the spin AM carried by the field. The complete characterization of the polarization requires the use of all the Pauli matrices $\hat{\vec{\sigma}} = \left(\hat{\sigma}_1, \hat{\sigma}_2, \hat{\sigma}_3\right)$, where $\hat{\sigma}_1$ and $\hat{\sigma}_2$ are unrelated to the AM [103]. A uniform polarization state is completely characterized by the expectation values

$$\vec{\mathfrak{S}} = \left(\tilde{\mathbf{e}}\middle|\hat{\vec{\sigma}}\middle|\tilde{\mathbf{e}}\right)_{\perp C}, \tag{2.24}$$

which form the normalized *Stokes vector* [104,105]. It can be considered as a *pseudo-spin* which represents the $\text{SU}\left(2\right)$ polarization state on the abstract $\text{SO}\left(3\right)$ Poincaré sphere – an analogue of the Bloch sphere. The pure helicity states $\left|\mathbf{e}^{\sigma}\right\rangle_{\perp}$ correspond to the poles of the Poincaré sphere, and the third component of the Stokes vectors determines the mean helicity: $\mathfrak{S}_3 = \bar{\sigma}$. Note also that in the basis of linear polarizations, the Stokes-vector components are determined by Eq. (2.24) with the permuted Pauli matrices $\hat{\vec{\sigma}}' = \left(\hat{\sigma}_3, \hat{\sigma}_1, \hat{\sigma}_2\right)$:

$$\vec{\mathfrak{S}} = \left(\tilde{\mathbf{e}}\middle|\hat{\vec{\sigma}}'\middle|\tilde{\mathbf{e}}\right)_{\perp L}. \tag{2.24'}$$

# 3. Polarization transformations and beam shifts at a dielectric interface

## 3.1. Snell-Fresnel reflection and transmission

In this Section we examine reflection and transmission of polarized Gaussian-type beams without intrinsic orbital AM, i.e., without vortex: $\ell = 0$. The geometry of the problem is depicted in Figures 1 and 4. A paraxial optical beam propagates in the $\left(x,z\right)$ plane at an angle $\vartheta$ to the $z$-axis and undergoes partial reflection and refraction at the plane interface $z = 0$ separating two non-absorbing dielectric media. The interface is characterized by the relative refractive index of the second medium, $n$, and its relative impedance $\zeta$.

In the *geometrical-optics* description, the beams are associated with their *central plane waves* which have wave vectors $\mathbf{k}_{\text{c}}^a$. Hereafter index $a = i, r, t$ denotes incident, reflected, and transmitted beams, respectively, and the index $i$ is omitted in the explicit expressions: $\mathbf{k}_{\text{c}}^i \equiv \mathbf{k}_{\text{c}}$, etc. The wave numbers in the three beams are $k^r = k^i \equiv k$ and $k^t = nk$. From Snell's law (which represents conservation of the tangent momentum components), it follows that the central wave vectors of all the three beams lie in the same $\left(x,z\right)$ plane: $\mathbf{k}_{\text{c}}^a = k^a\left(\sin\vartheta^a, 0, \cos\vartheta^a\right)$ and form angles [1]

$$\vartheta^i = \vartheta, \quad \vartheta^r = \pi - \vartheta, \quad \vartheta^t = \sin^{-1}\left(n^{-1}\sin\vartheta\right) \equiv \vartheta', \tag{3.1}$$

with the $z$-axis. It is natural to introduce coordinate frames of individual beams $\left(X^a, y, Z^a\right)$ with the $Z^a$-axes attached to their directions of propagation: $\mathbf{k}_{\text{c}}^a = k^a \mathbf{u}_Z^a$ (see Fig. 4). The beam frames are obtained via rotation of the laboratory coordinate frame $(x, y, z)$ by the angle $\vartheta^a$ about the $y$-axis:

$$\begin{pmatrix} \mathbf{u}_X^a \\ \mathbf{u}_y \\ \mathbf{u}_Z^a \end{pmatrix} = \hat{R}_y\left(\vartheta^a\right) \begin{pmatrix} \mathbf{u}_x \\ \mathbf{u}_y \\ \mathbf{u}_z \end{pmatrix}, \quad \hat{R}_y\left(\vartheta\right) = \begin{pmatrix} \cos\vartheta & 0 & -\sin\vartheta \\ 0 & 1 & 0 \\ \sin\vartheta & 0 & \cos\vartheta \end{pmatrix}, \tag{3.2}$$

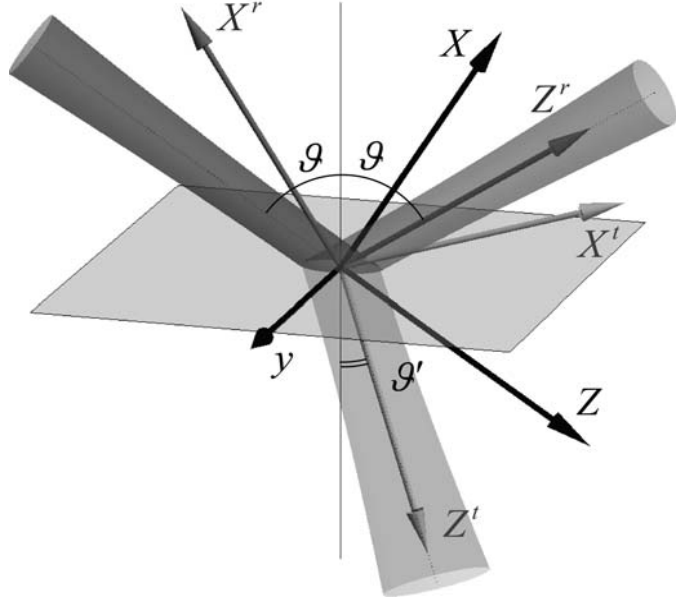

**Fig. 4.** Geometrical-optics scheme (without shifts) of reflection and transmission of a paraxial beam at a plane dielectric interface ($z = 0$). The plane of incidence of the beam is $(x, z)$. The beam coordinate frames $\left(X^a, y, Z^a\right)$ attached to the incident, reflected, and transmitted beams are shown. The laboratory frame is associated with the interface $z = 0$ and plane of incidence $y = 0$.

The electric field of the incident central plane wave can be written as $\mathbf{E}_\perp = e_X \mathbf{u}_X + e_y \mathbf{u}_y$, where we assumed normalization $|\mathbf{E}_\perp|^2 = |e_X|^2 + |e_y|^2 = 1$. The field components form the effective normalized *Jones vector* in the beam frame, $|\mathbf{E}\rangle_{\perp B} = |\mathbf{e}\rangle_{\perp B} = \left(e_X, e_y\right)^T$, and for the central waves the $X$-$y$ basis coincides with the basis of TM ($p$) and TE ($s$) modes with respect to the interface. The fields of the secondary beams are determined by the Fresnel reflection and transmission coefficients, $R_{p,s}\left(\vartheta\right)$ and $T_{p,s}\left(\vartheta\right)$ for the $p$ and $s$ modes [1]. Namely, implying natural transitions between the corresponding beam coordinate frames $\left(\mathbf{u}_X^a, \mathbf{u}_y, \mathbf{u}_Z^a\right)$, the transverse electric fields of the secondary beams are obtained by application of the effective *Fresnel Jones matrix* $\hat{F}^a$:

$$|\mathbf{E}^a\rangle_{\perp B} = \hat{F}^a |\mathbf{E}\rangle_{\perp B}, \quad \hat{F}^a = \begin{pmatrix} f_p^a & 0 \\ 0 & f_s^a \end{pmatrix}, \tag{3.3}$$

where we denoted $f^{i,r,t} \equiv 1, R, T$. The normalized Jones vectors for the beams take the form:

$$|\mathbf{e}^a\rangle_{\perp B} = Q^{a-1} |\mathbf{E}^a\rangle_{\perp B}, \quad Q^{a 2} = \langle \mathbf{E}^a | \mathbf{E}^a\rangle_{\perp B}. \tag{3.4}$$

Here $Q^a = \sqrt{\left|f_p^a\right|^2 |e_X|^2 + \left|f_s^a\right|^2 |e_y|^2}$ are the amplitude coefficients of the fields, whereas their energy coefficients, satisfying the conservation law, are given by [1]:

$$Q^a = \zeta^a \left| \frac{\cos \vartheta^a}{\cos \vartheta} \right| Q^{a2} , \text{ where } Q^r + Q^t = Q = 1 . \tag{3.5}$$

Here $\zeta^a$ is the relative impedance of the medium, i.e., $\zeta^i = \zeta^r = 1$ and $\zeta^t = \zeta$ .

Equations (3.1)–(3.5) describe the Snell-Fresnel reflection and transmission of the central plane wave in the beam.

## 3.2. Beam field transformation

Now, let us examine real beams possessing *finite Fourier spectra* with wave vectors $\mathbf{k}^a$ narrowly distributed around the central wave vectors $\mathbf{k}_c^a$. For a monochromatic field, $|\mathbf{k}^a| = |\mathbf{k}_c^a| = k^a$, and the wave vectors can be characterized by small orthogonal deflections $\boldsymbol{\chi}^a$:

$$\mathbf{k}^a = \mathbf{k}_c^a + \boldsymbol{\chi}^a , \quad |\boldsymbol{\chi}^a| \ll k^a , \quad \boldsymbol{\chi}^a \simeq k^a \left( \mu^a \mathbf{u}_X^a + \nu^a \mathbf{u}_y \right) . \tag{3.6}$$

Thus, $\mu$ and $\nu$ specify the in-plane and out-of-plane deflections of non-central wave vectors, as shown in Fig. 5. First, the Snell's law provides a connection between the components of the wave vectors (3.6):

$$\mu^a = \frac{k}{k^a} \gamma^a \mu , \quad \nu^a = \frac{k}{k^a} \nu , \quad \gamma^a = \frac{\cos \vartheta}{\cos \vartheta^a} . \tag{3.7}$$

Here we introduced the coefficients of the *elliptical deformations* of the beams: $\gamma^i = 1$, $\gamma^r = -1$, and $\gamma^t = \cos \vartheta / \cos \vartheta^t$. Second, the field components are related by the Fresnel coefficients which are written for TM and TE plane waves, i.e., waves with the electric field being parallel and orthogonal to their planes of incidence. However, the plane of incidence of a wave with $\nu \neq 0$ does *not* coincide with the $(x, z)$ plane, and, hence, the basis of $p$ and $s$ modes differs from the natural beam $X$-$y$ basis (see Fig. 5). Indeed, the TM and TE modes with respect to the interface $z = 0$ are associated with polar and azimuthal polarizations along the basic vectors $\mathbf{u}_\theta$ and $\mathbf{u}_\phi$ of the global *spherical coordinate frame*, Eq. (2.8) and (2.9) (cf. [74]). Thus, the deceptively simple problem of the reflection or refraction of a wave beam essentially involves three coordinate frames for its adequate description: (i) the laboratory frame, (ii) the beam frames, and (iii) the spherical frame of the TM and TE modes.

Let the beam spectra be given in the $\left( X^a, y, Z^a \right)$ coordinates as $\left| \tilde{\mathbf{E}}^a \left( \mathbf{k}^a \right) \right\rangle_B$. Transition to the laboratory frame $(x, y, z)$ is realized by inverse rotations (3.2) $\hat{R}_y \left( -\vartheta^a \right)$, $\left| \tilde{\mathbf{E}}^a \right\rangle_L = \hat{R}_y \left( -\vartheta^a \right) \left| \tilde{\mathbf{E}}^a \right\rangle_B$, whereas the next transitions to the global spherical coordinates are accomplished with the help of the transformation $\hat{U}_S \left( \mathbf{k}^a \right) = \hat{R}_y \left( \theta^a \right) \hat{R}_z \left( \phi^a \right)$, Eq. (2.9): $\left| \tilde{\mathbf{E}}^a \right\rangle_S = \hat{U}_S \left( \mathbf{k}^a \right) \left| \tilde{\mathbf{E}}^a \right\rangle_L$. Here the spherical angles $\left( \theta^a, \phi^a \right)$ determine the direction of $\mathbf{k}^a$. The resulting rotational transformation to the basis of $p$ and $s$ modes is

$$\left| \tilde{\mathbf{E}}^a \right\rangle_S = \hat{U}_\vartheta \left( \vartheta^a, \mathbf{k}^a \right) \left| \tilde{\mathbf{E}}^a \left( \mathbf{k}^a \right) \right\rangle_B , \quad \hat{U}_\vartheta \left( \vartheta^a, \mathbf{k}^a \right) = \hat{R}_y \left( \theta^a \right) \hat{R}_z \left( \phi^a \right) \hat{R}_y \left( -\vartheta^a \right) . \tag{3.8}$$

For the central plane wave $\mathbf{k}^a = \mathbf{k}_c^a$, we have $\left( \theta^a, \phi^a \right) = \left( \vartheta^a, 0 \right)$ and $\hat{U}_\vartheta \left( \vartheta^a, \mathbf{k}_c^a \right) = \hat{1}$. For a non-central wave (3.6), small wave-vector deflections $\mu^a$ and $\nu^a$ induce changes in $\theta^a$ and $\phi^a$, respectively:

$$\theta^a \simeq \vartheta^a + \mu^a , \quad \phi^a \simeq \frac{\nu^a}{\sin \vartheta^a} . \tag{3.9}$$

Thus, the in-plane deflection $\mu$ changes the *angle of incidence*, whereas the out-of-plane deflection $\nu$ turns the *plane of incidence* by the angle $\nu/\sin\vartheta$ about the $z$ axis (Fig. 5). Substituting Eq. (3.9) into Eq. (3.8), we obtain the rotation matrix $\hat{U}_\vartheta$ in the linear approximation in $\boldsymbol{\chi}^a$:

$$\hat{U}_\vartheta\left(\vartheta^a,\boldsymbol{\chi}^a\right)\simeq\hat{R}_y\left(\vartheta^a+\mu^a\right)\hat{R}_z\left(\nu^a/\sin\vartheta^a\right)\hat{R}_y\left(-\vartheta^a\right)\simeq\begin{pmatrix}1 & \nu^a\cot\vartheta^a & -\mu^a \\ -\nu^a\cot\vartheta^a & 1 & -\nu^a \\ \mu^a & \nu^a & 1\end{pmatrix}. \quad (3.10)$$

In fact, only the transverse components of the fields are essential for the paraxial problem under consideration, which are described by the $2\times2$ upper left sector of Eq. (3.10). In the basis of linear and circular polarizations it yields:

$$\hat{U}_{\vartheta\perp}\simeq\begin{pmatrix}1 & -\Phi_B^a \\ \Phi_B^a & 1\end{pmatrix},\quad\hat{V}_\perp\hat{U}_{\vartheta\perp}\hat{V}_\perp^\dagger\simeq\begin{pmatrix}\exp\left(-i\Phi_B^a\right) & 0 \\ 0 & \exp\left(i\Phi_B^a\right)\end{pmatrix}=\exp\left(-i\hat{\sigma}_3\Phi_B^a\right), \quad (3.11)$$

Here the basis of circular polarizations corresponds to the helicity basis (2.10), and

$$\Phi_B^a=-\nu^a\cot\vartheta^a=-\phi^a\cos\vartheta^a \quad (3.12)$$

is the *Berry geometric phase,* Eqs. (2.13)–(2.15), induced by the azimuthal rotation of the plane of incidence by the angle $\phi^a$, Eq. (3.9). The geometric-phase matrix (3.11) represents the effective *Jones matrix of the spin-orbit interaction* caused by the transition from the beam coordinate frame to the global spherical frame. It is this free-space rotational transformation that results in the transverse IF shifts, i.e., the spin Hall effect of light.

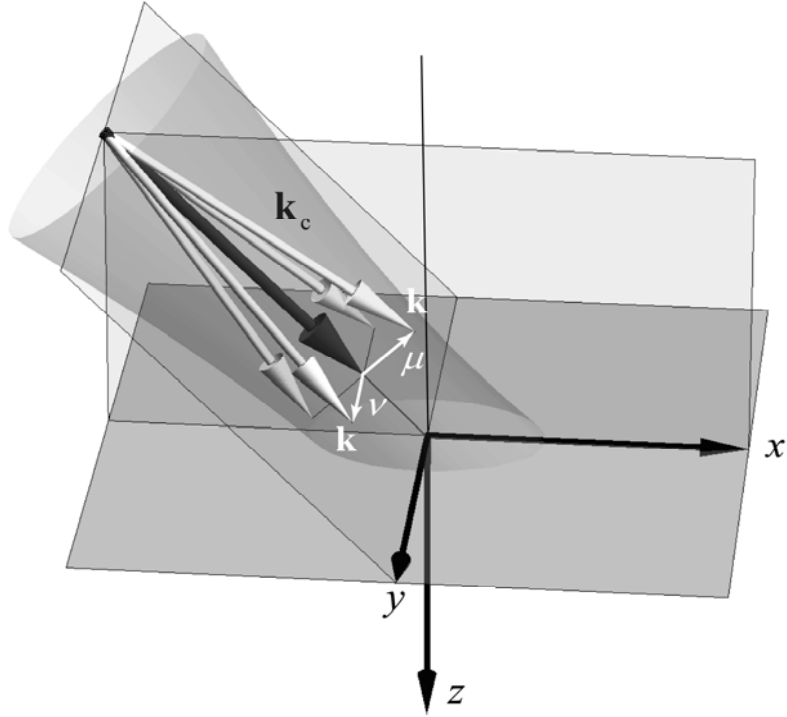

**Fig. 5.** Central wave vector $\mathbf{k}_c$ (black arrow) and non-central wave vectors $\mathbf{k}$ (white arrows) in the incident beam. Small in-plane and out-of-plane deflections, $\mu$ and $\nu$, vary the angle of incidence and the plane of incidence, respectively [see Eq. (3.6) and (3.9)].

After transforming the fields to the global spherical basis of *p* and *s* modes, one can connect them via the Fresnel boundary conditions (3.3):

$$\left|\tilde{\mathbf{E}}^a\right\rangle_{\perp S} = \hat{F}^a \left|\tilde{\mathbf{E}}\right\rangle_{\perp S}, \quad \hat{F}^a \simeq \begin{pmatrix} f_p^a\left(1 + \mu\dfrac{\partial \ln f_p^a}{\partial \vartheta}\right) & 0 \\ 0 & f_s^a\left(1 + \mu\dfrac{\partial \ln f_s^a}{\partial \vartheta}\right) \end{pmatrix}. \tag{3.13}$$

Here we took into account variations in the angle of incidence, Eq. (3.9): $f\left(\vartheta + \mu\right) \simeq f\left(\vartheta\right) + \mu f'\left(\vartheta\right)$. Corrections from the gradients of the Fresnel coefficients in Eq. (3.13) are responsible for the in-plane GH shifts of the beams.

Together, equations (3.6)–(3.13) yield the complete transformation and Jones matrix related the incident and secondary fields in the beam coordinate frames: $\hat{T}_\perp^a = \hat{U}_{\vartheta\perp}^\dagger\left(\vartheta^a, \chi^a\right)\hat{F}^a\left(\vartheta, \chi\right)\hat{U}_{\vartheta\perp}\left(\vartheta, \chi\right)$,

$$\left|\tilde{\mathbf{E}}^a\right\rangle_{\perp B} = \hat{T}_\perp^a \left|\tilde{\mathbf{E}}\right\rangle_{\perp B}, \quad \hat{T}_\perp^a \simeq \begin{pmatrix} f_p^a\left(1 + \mu\mathcal{X}_p^a\right) & f_p^a \nu \mathcal{Y}_p^a \\ -f_s^a \nu \mathcal{Y}_s^a & f_s^a\left(1 + \mu\mathcal{X}_s^a\right) \end{pmatrix}. \tag{3.14}$$

Here we introduced the quantities

$$\mathcal{X}_{p,s}^a = \frac{\partial \ln f_{p,s}^a}{\partial \vartheta}, \quad \mathcal{Y}_{p,s}^a = \left(1 - \gamma^{a-1}\frac{f_{s,p}^a}{f_{p,s}^a}\right)\cot \vartheta. \tag{3.15}$$

Equations (3.14) and (3.15) are the central results in this work; they describe field transformations upon the reflection and refraction of a paraxial optical beam at a plane dielectric interface [52,54,56,70,73]. The Jones matrix $\hat{T}_\perp^a$ characterizes effective *spatial dispersion* of the interface through $\mu$- and $\nu$-dependent terms. Importantly, the beam interaction with more complex interfaces (metallic, multi-layer, etc.) can be described by the same equations with the corresponding transmission and reflection coefficients. While the 'spin-orbit transformation' (3.11) is diagonal in the basis of circular polarizations, the Fresnel boundary conditions (3.13) are naturally diagonal in the basis of linear polarizations. Therefore, there is *no* global polarization basis that would diagonalize the whole transformation (3.14).

### 3.3. Gaussian beam shifts

To calculate the shifts explicitly, let us take an incident Gaussian beam at its waist $Z = 0$: $\left|\tilde{\mathbf{E}}\right\rangle_{\perp B} = \left|\mathbf{e}\right\rangle_{\perp B} G\left(\mu, \nu\right)$, where $\left|\mathbf{e}\right\rangle_{\perp B} = \left(e_x, e_y\right)^T$ is the Jones vector of the central plane wave and $G\left(\mu, \nu\right) = \dfrac{w_0^2}{2\pi}\exp\left[-\left(kw_0\right)^2\dfrac{\mu^2 + \nu^2}{4}\right]$ is the normalized Gaussian envelope with the waist $w_0$. By applying the Jones matrix (3.14) we obtain the fields of the secondary beams and calculate the expectation values of their transverse momenta and coordinates using Eqs. (2.18):

$$\left\langle X^a, Y^a\right\rangle = Q^{a-2}\left\langle \tilde{\mathbf{E}}^a\left|i\frac{\partial}{\partial k_{X,y}^a}\right|\tilde{\mathbf{E}}^a\right\rangle_{B\perp}, \quad \left\langle P_{X,y}^a\right\rangle = Q^{a-2}\left\langle \tilde{\mathbf{E}}^a\left|k_{X,y}^a\right|\tilde{\mathbf{E}}^a\right\rangle_{B\perp}. \tag{3.16}$$

Here the integration is taken over the transverse momentum components $k_X^a = \gamma^a k\mu$ and $k_y^a = k\nu$, and we took into account that the norms are equal to $N^a = \left\langle \tilde{\mathbf{E}}_\perp^a\left|\tilde{\mathbf{E}}_\perp^a\right\rangle\right. = Q^{a2}$.

In the case of *partial* reflection and transmission, $\sin \vartheta < n$, the Fresnel coefficients $f_{p,s}^a$ are real, and equations (3.14)–(3.16) at $Z^a = 0$ result in

$$\left\langle X^a\right\rangle = 0, \quad \left\langle P_X^a\right\rangle = \frac{\gamma^a}{k w_0^2}\frac{d \ln Q^a}{d\vartheta}, \tag{3.17}$$

$$\left\langle Y^a \right\rangle = -\frac{\overline{\sigma}}{2k} \frac{f_p^{a^2} \boldsymbol{\mathcal{Y}}_p^a + f_s^{a^2} \boldsymbol{\mathcal{Y}}_s^a}{Q^{a^2}}, \quad \left\langle P_y^a \right\rangle = \frac{\overline{\rho}}{k w_0^2} \frac{f_p^{a^2} \boldsymbol{\mathcal{Y}}_p^a - f_s^{a^2} \boldsymbol{\mathcal{Y}}_s^a}{Q^{a^2}} \tag{3.18}$$

Here $\overline{\sigma} = \left( \mathbf{e} \middle| \hat{\sigma}_2 \middle| \mathbf{e} \right)_{\perp B} = 2\,\mathrm{Im}\left( e_x^* e_y \right)$ is the average helicity of the incident beam (the Stokes parameter $\mathfrak{S}_3$) and $\overline{\rho} = \left( \mathbf{e} \middle| \hat{\sigma}_1 \middle| \mathbf{e} \right)_{\perp B} = 2\,\mathrm{Re}\left( e_x^* e_y \right)$ is the degree of linear polarization inclined at $\pi/4$ angle (the Stokes parameter $\mathfrak{S}_2$), Eq. (2.24').

In the case of total internal reflection, $\sin \vartheta > n$, there is no transmitted propagating wave, $f_{p,s}^r = 0$, while the reflection coefficients are complex: $f_{p,s}^r = \exp\left( i\delta_{p,s} \right)$, with real phases $\delta_{p,s}$. In this case equations (3.14)–(3.16) bring about

$$\left\langle X^{\mathrm{tot}\,r} \right\rangle = \frac{1}{k}\left( \left| E_X \right|^2 \mathrm{Im}\,\mathcal{X}_p^r + \left| E_y \right|^2 \mathrm{Im}\,\mathcal{X}_s^r \right), \quad \left\langle P_X^{\mathrm{tot}\,r} \right\rangle = 0, \tag{3.19}$$

$$\left\langle Y^{\mathrm{tot}\,r} \right\rangle = -\frac{1}{2k}\left[ \overline{\sigma}\,\mathrm{Re}\left( \boldsymbol{\mathcal{Y}}_p^r + \boldsymbol{\mathcal{Y}}_s^r \right) + \overline{\rho}\,\mathrm{Im}\left( \boldsymbol{\mathcal{Y}}_p^r - \boldsymbol{\mathcal{Y}}_s^r \right) \right], \quad \left\langle P_y^{\mathrm{tot}\,r} \right\rangle = 0. \tag{3.20}$$

The centroid displacement $\left\langle X^a \right\rangle$ and momentum shift (deflection) $\left\langle P_X^a \right\rangle$ represent the *spatial and angular Goos–Hänchen shifts*, whereas displacement $\left\langle Y^a \right\rangle$ and deflection $\left\langle P_y^a \right\rangle$ – *spatial and angular Imbert–Fedorov shifts*, respectively, see Fig. 1. The angular shifts depend on the beam waist and vanish in the paraxial limit $w_0 \to \infty$, whereas the spatial shifts are independent of the beam profile. Nonetheless, both spatial and angular shifts have a common origin: the *in-plane momentum gradients of the Fresnel coefficients* $\mathcal{X}_{p,s}^a$ for the GH effects (3.17) and (3.19) and the *geometric spin-orbit terms* $\boldsymbol{\mathcal{Y}}_{p,s}^a$ for the IF effects (3.18) and (3.20). The latter can be associated with the *transverse momentum gradients of the Berry phases* across the beams: $\partial \Phi_B^a / \partial \nu^a = -\cot \vartheta^a$. The IF effects show divergence at $\vartheta \to 0$ because the approximate transition to the spherical coordinates, Eqs. (3.9) and (3.10), becomes singular at normal incidence; in reality, of course, the transverse shift vanishes at $\vartheta = 0$. Away from the beam waist, at $Z^a \neq 0$, the real-space shifts grow according to the angular deviation: $\left\langle X^a \right\rangle \left( Z^a \right) = \left\langle X^a \right\rangle + \left\langle P_X^a \right\rangle Z^a / k^a$, $\left\langle Y^a \right\rangle \left( Z^a \right) = \left\langle Y^a \right\rangle + \left\langle P_y^a \right\rangle Z^a / k^a$, Fig. 1, which is used to measure small deviation effects [14,53,54,56,68,69].

The widely known spatial GH shift $\left\langle X^a \right\rangle$ is caused by the angular gradient of the *phase* of complex reflection or refraction coefficients and appears, e.g, upon the total internal reflection or interaction with complex dispersive interfaces [2–5,15–28]. In turn, the angular GH shift $\left\langle P_X^a \right\rangle$ is caused by the angular gradients of the *amplitude* of the Fresnel coefficients and can be observed in the case of partial reflection and refraction with real coefficients [9–14]. Thus, both GH effects are intimately related to the *spatial dispersion* of the scattering coefficients. The eigenmodes of the GH shifts are $p$ and $s$ linearly polarized waves, but polarization is not crucial for the existence of these phenomena, as it can also occur for scalar waves.

The IF shifts arise essentially owing to the intrinsic polarization properties of light. In the partial reflection/transmission case, the eigenmodes of the spatial shift $\left\langle Y^a \right\rangle$ are circularly polarized waves $\overline{\sigma} = \pm 1$, the shift is proportional to the helicity of the incident wave, and, thus, represents the *spin Hall effect of light*, Fig. 6. This transverse shift was discussed and studied for many years, both theoretically and experimentally [6–8,29–53]. Strikingly, while the clear theoretical explanation of the GH effect was given by Artmann the next year after its discovery [3], the accurate formulas (3.18) for the IF shift were derived and verified experimentally only recently [51–54] (although formula equivalent to Eq. (3.20) appeared first in the early paper by Schilling [7]). This could be explained by the fact that while the GH shifts can be fully explained within the 2D geometry of the plane of incidence, the transverse IF effect essentially involves a 3D description and requires an accurate characterization of the polarization of the incident field [52]. The angular IF shift $\left\langle P_y^a \right\rangle$ was revealed only recently [52–56,63,64]. Despite the fact that the eigenmodes of this effect are linearly polarized waves with $\bar{\rho} = \pm 1$, it originates from the same geometric-phase terms as the spatial transverse shift.

### 3.4. Linear and angular momentum conservation

Importantly, the spatial IF shift is intimately related to the balance and *conservation of the AM of light* [40,41,50–52]. Indeed, the rotational symmetry of the medium about the $z$-axis implies that the $z$-component of the total AM must be conserved upon reflection and refraction. Using equation (2.22) for paraxial beams, the spin AM per photon in the three beams can be written as $\left\langle \mathbf{S}^a \right\rangle = \bar{\sigma}^a \mathbf{u}_Z^a$, $\left\langle S_z^a \right\rangle = \bar{\sigma}^a \cos \vartheta^a$. From the Fresnel boundary conditions (3.3) and (3.4), one can determine the mean helicities $\bar{\sigma}^a = \left( \mathbf{e}^a \middle| \hat{\sigma}_2 \middle| \mathbf{e}^a \right)_{\perp B}$ which yield the spin AM for the cases of partial reflection/transmission and for total internal reflection:

$$\left\langle S_z^a \right\rangle = \bar{\sigma} \cos \vartheta^a \frac{f_p^a f_s^a}{Q^{a2}}, \quad \left\langle S_z^{\mathrm{tot}\, r} \right\rangle = -\left( \bar{\sigma} \cos \delta + \bar{\rho} \sin \delta \right) \cos \vartheta, \tag{3.21}$$

where $\delta = \delta_s - \delta_p$. The transverse coordinate shifts $\left\langle Y^a \right\rangle$, Eqs. (3.18) and (3.20), together with the momentum component $\left\langle P_x^a \right\rangle \simeq k \sin \vartheta$ generate the extrinsic orbital AM (2.20):

$$\left\langle L_z^{\mathrm{ext}\, a} \right\rangle = \left[ \left\langle \mathbf{R}^a \right\rangle \times \left\langle \mathbf{P}^a \right\rangle \right]_z = -\left\langle Y^a \right\rangle k \sin \vartheta. \tag{3.22}$$

Since the number of photons in the beams is proportional to the energy $Q^a$, we write the conservation of the $z$-component of the total AM in the scattering as

$$Q^r \left[ \left\langle S_z^r \right\rangle - \left\langle Y^r \right\rangle k \sin \vartheta \right] + Q^t \left[ \left\langle S_z^t \right\rangle - \left\langle Y^t \right\rangle k \sin \vartheta \right] = \left\langle S_z \right\rangle. \tag{3.23}$$

Here we took into account that $\left\langle L_z^{\mathrm{ext}} \right\rangle = \left\langle Y \right\rangle = 0$ and $Q = 1$ for the incident beam. Importantly, this equation is not satisfied with $\left\langle Y^r \right\rangle = \left\langle Y^t \right\rangle = 0$. Changes in the spin AM (3.21) which occur upon Snell-Fresnel reflection/refraction of the wave beam must be compensated by the non-zero extrinsic orbital AM (3.22) of the secondary beams. Substituting Eqs. (3.18), (3.20), and (3.21) into (3.23) and using relations (3.1)–(3.7) and (3.15), one can verify that Eq. (3.23) is fulfilled identically. Thus, the nonzero transverse shifts of the reflected and refracted beams ensure the conservation of the AM in the problem. In the case of total internal reflection, there is no transmitted beam, $Q^t = 0$, $Q^r = 1$, and the shift can be derived *solely* from the AM conservation:

$$\left\langle Y^{\mathrm{tot}\, r} \right\rangle = \frac{\left\langle S_z^{\mathrm{tot}\, r} \right\rangle - \left\langle S_z \right\rangle}{k \sin \vartheta}, \tag{3.24}$$

which, together with Eq. (3.21), immediately yields Eq. (3.20). This shows that this IF shift is very robust and is practically independent of the shape and fine details of the incident beam field. In other cases, the AM conservation imposes a constraint on the beam shifts but do not fix their values.

In a similar manner, one can verify that the angular shifts obey a constraint following from the *conservation of the linear momentum* [63,106]. The balance of the tangential components of the beam momenta yields:

$$-Q^r \left\langle P_X^r \right\rangle \cos \vartheta + Q^t \left\langle P_X^t \right\rangle \cos \vartheta' = 0, \quad Q^r \left\langle P_y^r \right\rangle + Q^t \left\langle P_y^t \right\rangle = 0. \tag{3.25}$$

Here we took into account the fact that the $x$-component of the momentum is $\langle P_x^a \rangle = \langle P_X^a \rangle \cos \vartheta^a + \langle P_Z^a \rangle \sin \vartheta^a$ and the balance of $\langle P_Z^a \rangle \sin \vartheta^a = k^a \sin \vartheta^a = k \sin \vartheta$ is guaranteed by Snell's law (3.1). Using Eqs. (3.1)–(3.7) and (3.15), one can readily show that Eqs. (3.17) and (3.18) fulfil Eqs. (3.25) identically. Note, however, that the linear momentum conservation (3.25) could be satisfied without angular shifts, i.e., at $\langle P_{X,y}^r \rangle = \langle P_{X,y}^t \rangle = 0$ (which is achieved for special beam models [52]), while the AM balance (3.23) essentially requires at least one transverse shift, $\langle Y^r \rangle$ or $\langle Y^t \rangle$, to be nonzero. In the case of total internal reflection ($Q^t = 0$, $Q^r = 1$), the momentum conservation (3.25) results in vanishing of angular shifts, $\langle P_{X,y}^{\text{tot}\,r} \rangle = 0$, in agreement with (3.19) and (3.20).

Thus, the 'microscopic' plane-wave interference resulting in the shifts (3.17)–(3.20) and 'macroscopic' arguments of the conservation laws (3.23)–(3.25) are in complete agreement with each other. This demonstrates a remarkable interplay of the 'geometric' and 'dynamical' mechanisms in the wave interaction phenomena.

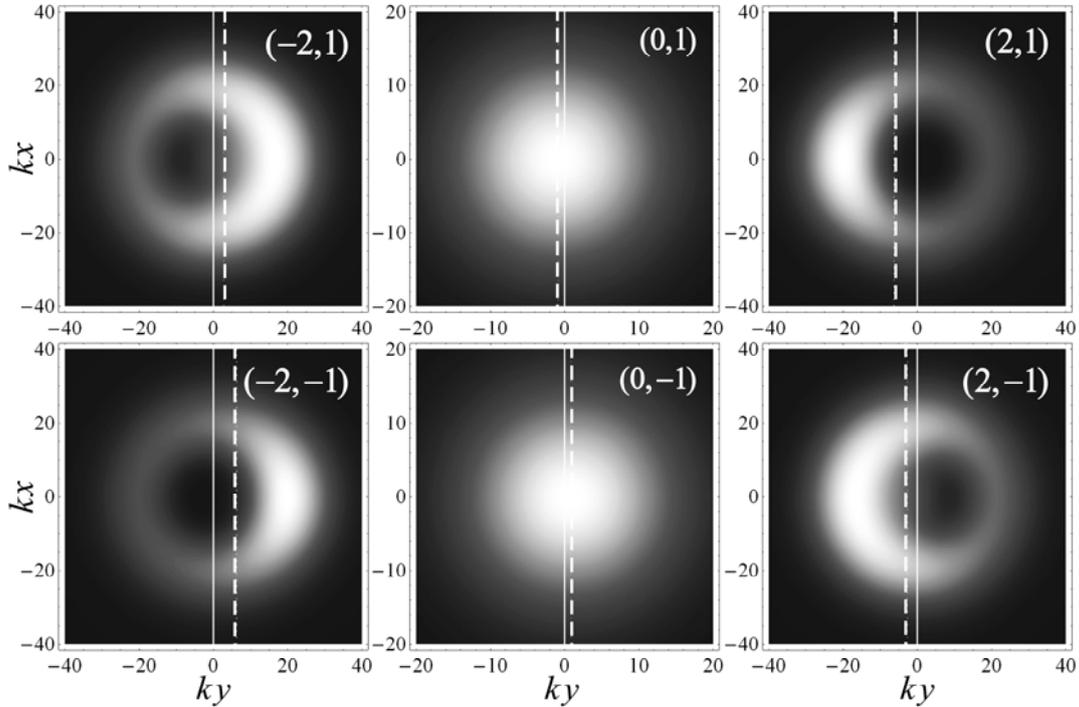

**Fig. 6.** Transverse intensity distributions (at $Z^r = 0$) in the Laguerre–Gaussian beams reflected from the air-glass interface ($n = 1.5$). The beams are marked by their orbital and spin AM quantum numbers $(\ell, \sigma)$ (the radial number $p = 0$). The angle of incidence $\vartheta = \pi / 3$, whereas the aperture angle is $\theta_a \equiv 2 / (k w_0) = 0.1$. The dashed lines indicate the IF shifts of the centroids given by Eqs. (3.18) and (4.4). The spin-dependent shifts of the Gaussian beam represent perfect translations of the whole circularly-polarized beam, while the vortex-beam shifts are accompanied by deformations of the intensity distributions.

# 4. Goos-Hänchen and Imbert-Fedorov shifts of vortex beams

Let us consider now the beam shift effects in the case of the incident vortex beam with $\ell \neq 0$, which carries intrinsic orbital AM. Similar to the Fresnel boundary conditions (3.3) and (3.4) that determine the changes in the spin AM at the boundary, Eqs. (3.21), one can determine the boundary conditions for the change of the intrinsic orbital AM at the interface. The orbital AM of the incident beam is $\langle \mathbf{L} \rangle = \ell \mathbf{u}_z$, Eq. (2.22). Its changes in the secondary beams arise from *elliptical deformations* of the beams, $\gamma^a$, Eq. (3.7). Using simple geometrical considerations and calculations, one can show that the intrinsic orbital AM of the elliptically deformed paraxial vortex beam acquires an additional factor of $\left( \gamma^a + \gamma^{a^{-1}} \right) / 2$ [62]. Thus, the intrinsic AM of the secondary beams can be written as [62,63]

$$\langle \mathbf{L}^{\mathrm{int}\, a} \rangle = \frac{\ell}{2} \left( \gamma^a + \frac{1}{\gamma^a} \right) \mathbf{u}_z^a, \ \text{ i.e., } \ \langle \mathbf{L}^{\mathrm{int}\, r} \rangle = -\ell \mathbf{u}_z^r, \ \langle \mathbf{L}^{\mathrm{int}\, t} \rangle = \frac{\ell}{2} \left( \frac{\cos \vartheta}{\cos \vartheta'} + \frac{\cos \vartheta'}{\cos \vartheta} \right) \mathbf{u}_z^t. \quad (4.1)$$

As we already know, changes in the $z$-components of the intrinsic AM must be compensated by transverse shifts producing extrinsic orbital AM (3.22). It turns out that the vortex-generated transverse shifts can be obtained separately from the AM conservation between incident and reflected or incident and transmitted beams. As a result, these shifts read [cf. Eq. (3.24)] [59,62,63]

$$\langle Y^a \rangle = \frac{\langle L_z^{\mathrm{int}\, a} \rangle - \langle L_z^{\mathrm{int}} \rangle}{k \sin \vartheta}, \ \text{ i.e., } \ \langle Y^r \rangle = 0, \ \langle Y^t \rangle = \frac{\ell}{2k} \tan \vartheta \left( 1 - n^{-2} \right), \quad (4.2)$$

and only the transmitted beam experiences this shift.

The $\ell$-dependent transverse shift (4.2) represents an example of the *orbital Hall effect* of light caused by the orbit-orbit interaction between the intrinsic and extrinsic parts of the orbital AM. Akin to the IF shifts (3.20) and (3.24), this shift is very robust with respect to the details of the incident beam field. Note that Eqs. (4.1) and (4.2) is independent of polarization and, hence, can be regarded as the shift of a *scalar* wave beam at a planar interface. The dependence on the angle of incidence $\vartheta$ is of purely geometrical origin and does not imply any spatial dispersion caused by the interface. Yet, the $\langle Y \rangle \propto k^{-1} \tan \vartheta$ dependence in Eq. (4.2) demonstrates drastic difference as compared to the IF shift (3.18) and (3.20), which behaves as $\langle Y \rangle \propto k^{-1} \cot \vartheta$.

In addition to the 'scalar' $\ell$-dependent Hall effect (4.2), there are also $\ell$-dependent phenomena caused by polarization properties of the vortex beams interacting with an interface. Specifically, the angular deviations of the GH and IF effects become coupled to the complex vortex structure and induce additional shifts that depend on both polarization and vortex charge. To show this, we note that the angular shifts in Eqs. (3.17) and (3.18) can be considered as *imaginary spatial shifts* in the Gaussian envelopes of the beams, so that the resulting complex shifts can be written as [54,55]:

$$\delta X^a = \langle X^a \rangle - i \gamma^{a-2} \frac{w_0^2}{2} \langle P_x^a \rangle, \quad \delta Y^a = \langle Y^a \rangle - i \frac{w_0^2}{2} \langle P_y^a \rangle. \quad (4.3)$$

At the same time, an imaginary shift in the vortex distribution induces the *orthogonal real shift*, proportional to the vortex charge. This is clearly seen if we write the vortex structure as $\left[ \gamma^a X^a + i \, \mathrm{sgn}(\ell) \, y \right]^{|\ell|}$. Obviously, here the imaginary shift along $X^a$ is equivalent to the real shift along $y$, which is magnified by the power $\ell$, and vice versa. After straightforward calculations, we obtain that the *angular* GH and IF shifts induce the *spatial* $\ell$-dependent shifts in the vortex beams (shown in Fig. 6) [63,64]:

$$\langle X^a \rangle = \ell \gamma^{a-1} \frac{w_0^2}{2} \langle P_y^a \rangle \Big|_{\ell=0}, \quad \langle Y^a \rangle = -\ell \gamma^{a-1} \frac{w_0^2}{2} \langle P_x^a \rangle \Big|_{\ell=0}. \quad (4.4)$$

In turn, the angular shifts are also modified by the vortex structure, but these changes might depend on the particular shape of the beam envelope [81]. In the case of the Laguerre–Gaussian incident beam, the vortex magnifies the angular shifts by a factor of $\left( 1 + |\ell| \right)$, so that the additional contribution is [63,64]:

$$\left\langle P_X^a \right\rangle = |\ell| \left\langle P_X^a \right\rangle \Big|_{\ell=0}, \quad \left\langle P_y^a \right\rangle = |\ell| \left\langle P_y^a \right\rangle \Big|_{\ell=0}. \tag{4.5}$$

The $\ell$-dependent shifts (4.4) and (4.5) were measured experimentally in [60,64]. They rather represent an interplay between the polarization-dependent GH or IF shifts and the complex vortex structure. These shifts do not affect the momentum and AM balance, because the original angular shifts already satisfied Eq. (3.25).

In total, the spatial and angular shifts for polarized vortex beams reflected and refracted at a dielectric interface are given by the sum of *three contributions*: (i) GH and IF shifts (3.17)–(3.20), (ii) the 'scalar' orbital Hall effect (4.2) related to the tilt of the transmitted beam, and (iii) additional shifts (4.4) and (4.5) due to the coupling between the polarization-dependent angular shifts and the complex vortex structure. The IF shifts (i) and (iii) are exemplified in Fig. 6 displaying the intensity distributions and centroids of the Laguerre–Gaussian beams reflected from the air-glass interface. Note that, while the spin-dependent shifts of the Gaussian beam manifest themselves in a perfect translations of paraxial beams, the vortex beam shifts are accompanied by significant *deformations* of the beam profiles [78,79,107]. It is worth noticing that there are no $\ell$-dependent shifts in the case of total internal reflection: the $z$-component of the intrinsic orbital AM is conserved due to the effective flip of $\ell$ in the reflected beam, Eq. (4.1), and there are no angular shifts to be coupled with the vortex, Eqs. (3.19) and (3.20).

# 5. Extensions and related problems

From the above analysis it follows that the GH and IF beam shifts represent basic phenomena which appear in any reflection and refraction process. We have considered only the simplest configuration of a uniformly polarized Gaussian-type and vortex beams and plane isotropic dielectric interface without losses. More complicated situations and various modifications of the GH and IF shifts have been widely considered during the past several years. Here we give a brief overview of the most important extensions and related beam-shift problems.

1. Weak measurements. The GH and IF shifts of the beam centroids have a typical magnitude $k^{-1}$, i.e., a fraction of the *wavelength*. Measurement of such shifts is a challenging problem for experimentalists. However, there is a method of "*quantum weak measurements*" [108–111] which allows significant magnification of the beam shifts and measurements of the elements of the interface Jones matrix (3.14). Roughly speaking, the method consists in the use of two *almost orthogonal* polarizers: One *pre-selecting* the incident beam state and the other *post-selecting* reflected or refracted beam states after the interface. Depending on the mutual orientation of the two polarizers, one can increase the shift of the post-selected beam up to the order of the *beam width*.

The weak-measurement technique was used in experiments [53,56,68,69,112–115] measuring the IF effect, whereas its detailed theoretical analysis in the beam-shift problem is given in [53,54,116,117]. In fact, the weakly measured beam shift yields the corresponding element of the interface Jones matrix (3.14) which is interpreted as an effective Hamiltonian of the polarization-momentum coupling. A predecessor of the weak-measurement technique was the idea to measure the splitting of the beam intensity in the crossed input and output polarizers [38,51].

2. Higher-order deformations. The shifts of the beam centroid are the first moments of the intensity distributions, so that they describe only the simplest *first*-order deformations of the beam (Fig. 1). At the same time, the reflected and transmitted beams can undergo fine *higher-order deformation (aberration)* effects, which are not accounted in the beam shifts (see, e.g., the vortex-beam deformations in Fig. 6). Such deformations can be described, e.g., via expanding the beam intensity as a superposition of the orbital AM vortex eigenmodes taken with respect to the geometrical-optics axis. In this manner, the shift and extrinsic orbital AM of the secondary beams arises from the superposition of vortex modes with $\ell$ differing by $\pm 1$. Calculations and measurements of the *orbital AM spectrum* of the reflected beam were presented recently [118]. In the case of the incident vortex beam, the shift effects are accompanied by the shift and fine splitting of the charge-$\ell$ vortex into a *constellation* of the $|\ell|$ unit strength vortices. Such constellation characterizes the beam aberrations and properties of the interface up to the $|\ell|$ order [119].

3. Special angles. The beam reflection from a plane dielectric interface has two peculiar points: the *Brewster angle* of incidence and the *critical angle* upon the reflection from a less dense medium. The beam shifts can be significantly enhanced in the vicinity of these angles (the spin Hall effect diverges at the Brewster angle, while the GH shift diverges at the critical incidence), and the problem of the beam reflection can be challenging for the accurate theoretical treatment [74]. The beam shifts are well studied, both theoretically and experimentally, in the vicinity of the Brewster angle [11,14,60,68,72,112,120,121]. At the same time, the situation is more complicated for the near-critical incidence, where the theoretical and experimental results are still far from good agreement. Indeed, see [122–124] for the GH effect, and we have found that the previously reported measurements of the IF shift near the critical angle [35,38,44,47,49] have values about 1.5 times larger than those predicted by the Schilling formulae (3.20). The plane-wave Fourier analysis of the beam reflection becomes quite complicated near the critical angle [74], and perhaps one should apply real-space boundary conditions at the interface to find the actual reflected beam transformations [125]. It is also possible that the transverse energy-flow effect in the evanescent transmitted field, which was originally discovered by Fedorov and Imbert [6,8], still contributes the IF shift near the critical angle.

4. Complex incident beams. The beam-shift effects can also be sensitive to the *shape* of the incident beam. In addition to the Gaussian and Laguerre-Gaussian beams, the beam shifts were considered for the following types of beams: paraxial Bessel [81], non-paraxial Bessel [126], Laguerre-Gaussian with higher radial order [127], Hermite-Gaussian [128,129], and for the dipole radiation [74]. General aspects of the shape-dependent and shape-independent beam shifts are analyzed in [130]. Another important characteristic of the incident beam is the degree of its *spatial coherence*. Dependence of different beam shifts on this characteristic caused a theoretical controversy [131–134] which was recently resolved by experimental measurements [135,136].

5. Complex media. Apparently, the most popular extensions of the beam-shift problems are generalizations of these effects to the cases of various *complex media*. These include metals [26,137,138], dissipative media [15,17,55,66,67,120,139,140], semiconductors [66,67], various anisotropic (including chiral) media [69,141–145], multilayered structures [48,146–150], metamaterials [23,70,141,147,151,152], photonic crystals [24,148,150,153], thin films [48,113,114], resonators [16,44,154], plasmonic systems [19,23,25,87,115,155], curved interfaces (lenses) [13,87,154,156–158], nonlinear media [18,67,159], etc. In addition, the GH shift is studied for matter waves in various quantum and condensed-matter systems [27,28,160].

6. Related effects. It is important to mention effects related to the beam shifts that occur upon reflection/refraction at sharp interfaces.

First, propagation of light in a *gradient-index medium* exhibits an AM-dependent transverse transport of the beam [43,50,61,161–167]. This includes the *spin and orbital Hall effects* (also referred to as optical Magnus effect [43,168]). The Hall effects in a gradient-index medium can be derived as a limiting case of the spatial IF shifts when the beam is transmitted through multiple dielectric interfaces with low contrasts [43,50,61]. This effect is also intimately related to the AM conservation, and is described by the action of the '*geometric Berry force*' in the momentum space. (Note that the IF shift at a sharp interface can be attributed to the action of the so-called 'Abraham force' related to the difference between the Abraham and Minkowski momenta of photons in a dielectric medium [169].)

Another related beam-shift effect is the *geometric Hall effect of light* [170–172], which occurs upon observation of a beam of light in a *tilted cross-section*. This effect can be observed even in free space via measurements of the energy flux density through the tilted cross-section of the beam.

The geometric-Hall beam shifts are also intimately related to the transformations of the AM and are similar to the orbital Hall-effect shift (4.2) – both phenomena have characteristic $\ell \tan \vartheta / 2k$ dependence, which originates from the spatial tilt rather than gradients in the momentum Fourier space. In the space-time domain, an entirely analogous beam shift occurs upon observation of the beam in the relativistic moving frame; this is the *relativistic Hall effect* [173].

# 6. Conclusion

We have examined the beam interaction with a plane dielectric interface. To give a self-consistent tutorial description of the beam shifts, we have introduced the basic concepts of the coordinate rotations, geometric phases, and angular momentum. The wave field transformations at the interface have been described based on the Snell-Fresnel reflection and transmission formulas as applied to constituent plane waves in the beam spectrum. This provides an effective Jones matrix of the interface in the momentum representation, which describes all beam deformations in the first post-paraxial approximation. The matrix possesses effective spatial dispersion and contains two momentum-dependent terms: (i) the one arising from the real spatial dispersion of the Fresnel coefficients and (ii) another one appearing as a result of the geometric phase difference between constituent plane waves propagating in different planes. The first term is responsible for in-plane beam deformations and GH shifts, whereas the second term gives rise to the out-of-plane IF shift, i.e., spin Hall effect of light. Importantly, while the GH terms in the Jones matrix are diagonal in the basis of linear polarizations, the IF term becomes diagonal in the basis of circular polarizations. Since both terms are momentum-dependent, there are no global polarization eigenmodes, so that the interface always produces polarization mixing in the beam.

Calculating the reflection and transmitted beams from the effective Jones matrix, one can readily determine coordinates of their centroids. This immediately yields expressions for the spatial and angular GH and IF shifts. We have considered these shifts for both partial and total internal reflections, and shown that they satisfy the conservation laws for the normal component of the total AM and tangential component of the linear momentum. Thus, if the geometric phases underlie the IF shift on the *local* level of constituent plane waves, the conservation of the AM underpins this effect on the *global* level of the integral beam characteristics. Note that while post-paraxial terms are essential for constituent plane waves, the angular and linear momentum can be considered in the paraxial limit. It should be emphasized that the presented approach is rather universal and can be easily generalized to various types of interfaces – including dissipative, metal, anisotropic, and layered structures.

In addition to the polarization transformations of the Gaussian-type beams, we have also considered reflection and refraction of vortex beams carrying intrinsic orbital AM. Using conservation of the AM and properties of optical vortices, we have shown that the secondary beams experience vortex-dependent spatial and angular GH and IF shifts. Being proportional to the vortex charge, these shifts can be significantly enhanced for high-order beams. Numerical calculations demonstrated that while the spin-dependent IF shift of the circularly-polarized Gaussian beam represents a perfect translation, the transverse shifts of vortex beams are accompanied by beam deformations.

Finally, we have given an overview of the most important extensions and generalizations of the basic GH and IF effects. These include quantum weak measurement approach, higher-order aberration effects, special Brewster and critical angles of incidence, complex shapes of the incident beam, complex media or interfaces, and related beam-shift effects in other problems.

We are indebted to M.A. Alonso, M.R. Dennis, V.G. Fedoseyev, and Y. Li for fruitful discussions and relevant remarks. This work was supported by the European Commission (Marie Curie Action), Science Foundation Ireland (Grant No. 07/IN.1/I906), and von Humboldt Foundation.